\documentclass[sigconf]{acmart}
\AtBeginDocument{%
  }

\usepackage{times}
\usepackage{soul}
\usepackage{url}
\usepackage{multirow} 
\usepackage{enumitem}
\usepackage{picinpar}
\usepackage{bm}
\usepackage[normalem]{ulem}
\usepackage{subcaption}
\usepackage{xcolor}  
\newtheorem{proposition}{Proposition}

\setcopyright{acmlicensed}
\copyrightyear{2018}
\acmYear{2018}
\acmDOI{XXXXXXX.XXXXXXX}
\acmConference[Conference acronym 'XX]{Make sure to enter the correct
  conference title from your rights confirmation email}{June 03--05,
  2018}{Woodstock, NY}
\acmISBN{978-1-4503-XXXX-X/2018/06}

\begin{document}

\title{CityLight: A Neighborhood-inclusive Universal Model for Coordinated City-scale Traffic Signal Control}

\newcommand{\thuaffil}{%
  \affiliation{%
    \institution{Department of Electronic Engineering, Tsinghua University}
    \city{Beijing}
    \country{China}
  }
}
\author{Jinwei Zeng}
\thuaffil

\author{Chao Yu}
\thuaffil

\author{Xinyi Yang}
\thuaffil

\author{Wenxuan Ao}
\thuaffil

\author{Qianyue Hao}
\thuaffil

\author{Jian Yuan}
\thuaffil

\author{Yong Li}
\thuaffil

\author{Yu Wang}
\thuaffil

\author{Huazhong Yang}
\thuaffil

\renewcommand{\shortauthors}{Trovato et al.}

\begin{abstract}
City-scale traffic signal control (TSC) involves thousands of heterogeneous intersections with varying topologies, making cooperative decision-making across intersections particularly challenging. Given the prohibitive computational cost of learning individual policies for each intersection, some researchers explore learning a universal policy to control each intersection in a decentralized manner, where the key challenge is to construct a universal representation method for heterogeneous intersections. However, existing methods are limited to universally representing information of heterogeneous ego intersections, neglecting the essential representation of influence from their heterogeneous neighbors. Universally incorporating neighborhood information is nontrivial due to the intrinsic complexity of traffic flow interactions, as well as the challenge of modeling collective influences from neighbor intersections. To address these challenges, we propose CityLight, which learns a universal policy based on representations obtained with two major modules: a Neighbor Influence Encoder to explicitly model neighbor's influence with specified traffic flow relation and connectivity to the ego intersection; a Neighbor Influence Aggregator to attentively aggregate the influence of neighbors based on their mutual competitive relations. Extensive experiments on five city-scale datasets, ranging from 97 to 13,952 intersections, confirm the efficacy of CityLight, with an average throughput improvement of 11.68\% and a lift of 22.59\% for generalization. Our codes and datasets are released: \url{https://github.com/JinweiZzz/CityLight}.  
\end{abstract}




\maketitle

\section{Introduction} 
Given the severe traffic congestion in modern metropolitan cities~\cite{afrin2020survey}, developing effective traffic signal control (TSC) methods for city-scale traffic management becomes imperative. With the goal of ensuring the smooth passing of vehicles inside and between intersections, TSC methods are expected to greatly improve traffic efficiency~\cite{chien2006cost}, reduce congestion~\cite{wu2018distributed}, and decrease vehicle emissions~\cite{zhang2009assessing}.   

Despite great efforts in TSC, cooperative decision-making across thousands of heterogeneous intersections with varying topologies and arbitrary scales is still at a preliminary stage and has shortcomings. Conventional methods~\cite{koonce2008traffic,varaiya2013max} rely on human-crafted decision rules and lack flexibility in addressing complex traffic dynamics. Multi-agent reinforcement learning (MARL) methods excel in flexibility by adaptively learning optimal strategies based on environmental feedback, but the city-scale optimization urges the construction of a universal policy for resource and training efficiency considerations. However, since universal policy methods assign a shared policy network for heterogeneous intersections, a key point here is to universally represent these intersections with varying topologies and physical properties so that the shared policy can effectively interpret and generalize across diverse intersections. Nevertheless, existing universal policy methods are limited to achieving the representation of heterogeneous ego intersections' information~\cite{oroojlooy2020attendlight,liang2022oam}. The effective universal representation of neighbor intersection influences is insufficiently addressed, leaving space for performance improvement.  


\begin{figure}[t]
    \centering
    \includegraphics[width=0.99\columnwidth]{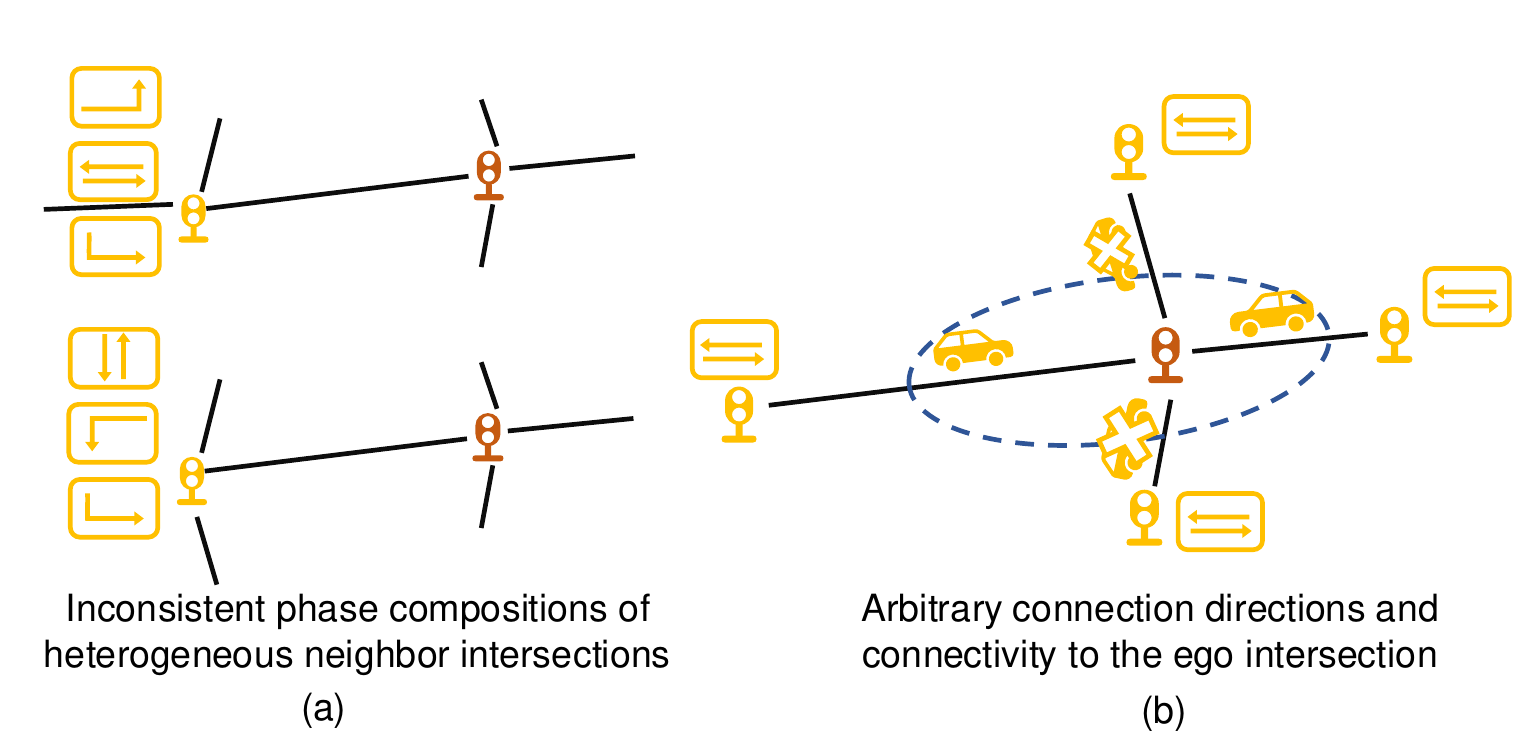}
    \caption{Diverse and intricate neighbor intersection traffic flow relations come from (a) heterogeneity of neighbor intersections that causes inconsistent phase compositions; (b) arbitrary connecting directions and connectivity from neighbor intersections to ego intersection.}
    \label{fig:background}
    \Description{background}
\end{figure}

\textbf{Universally representing and aggregating neighborhood influences} to enhance cooperation in the learned universal policy is non-trivial and challenging. Firstly, heterogeneous neighbor intersections have inconsistent traffic flow relations to the ego intersection (Fig~\ref{fig:background}(a)), and the diversity is even exacerbated by their arbitrary connecting directions and connectivity (indicated by distance and number of connecting lanes) (Fig~\ref{fig:background}(b)). Representing neighbor intersections' influences under the varying traffic flow relation and connectivity in an aligned measure is challenging. Secondly, the rotational symmetry of intersections governs the inherent competitive and collaborative relationships between neighboring intersections, increasing the complexity of modeling their combined influences when aggregating their representations.

To address this challenge, we propose \textbf{CityLight}, a MARL-based method to learn a universal policy based on representations that consider the influences of heterogeneous neighbor intersections. Built on the parameter-sharing Multi-Agent Proximal Policy Optimization (MAPPO) algorithm~\cite{yu2022surprising}, CityLight comprises two major modules to universally representing the intricate neighbor influences: a Neighbor Influence Encoder to project neighbor observations onto a uniform influence space based on specified traffic relations and connectivities to the ego intersection; a Neighbor Influence Aggregator to attentively fuse neighbor intersections' influence representations based on their mutual collaborative and competitive relations. We further introduce a neighborhood-inclusive reward design to better incorporate neighborhood outcomes into the optimization process. Extensive experiments on five datasets that comprise 97 to 13952 heterogeneous real-world intersections show that CityLight consistently outperforms both state-of-the-art universal policy models and individual policy methods, bringing an average 11.68\% lift of the throughput. Meanwhile, CityLight shows superior generalizability, achieving a 22.59\% increase of throughput in transfer scenarios. These comprehensive evaluations validate our designs' strength in effectively aligning and aggregating diverse and intricate neighborhood influences, enabling the universal policy to learn the shared strategy of neighborhood coordination across heterogeneous intersections. Our main contributions can be summarized as: 


\begin{itemize}
    \item We target the challenging city-scale traffic signal control problems, extending existing universal policy methods from universally representing only ego intersections' information to also effectively representing heterogeneous neighbors' influences.
    \item By proposing a Neighbor Influence Encoder and a Neighbor Influence Aggregator that project neighbor observations into a uniform influence space and attentively fuse their representations based on traffic relations and interactions with the ego intersection, CityLight manages to learn a cooperative universal policy for city-scale heterogeneous intersections.
    \item Extensive experiments on five city-scale datasets with up to 14k intersections validate the effectiveness of CityLight, with an overall 11.68\% improvement and a superiority of 22.59\% in transfer scenarios in throughput. 
    \item We have open-sourced our high-authenticity city-scale datasets, which are the first to achieve the scale of 10k intersections, with the hope of facilitating research advancement in city-scale traffic signal control.
\end{itemize}

\section{Related Work}
City-scale traffic signal control is a vital measure for traffic efficiency enhancement, which has therefore been extensively researched. \textbf{Conventional} approaches generally incorporate manually designed rules for signal transition judgement~\cite{martinez2011survey,roess2004traffic,cools2013self,koonce2008traffic}. Although applicable to large-scale intersections, rule-based methods lack flexibility in addressing complex traffic dynamics and achieving coordination~\cite{friesen2022multi}, resulting in limited effectiveness. Recent advancements in reinforcement learning have led to a surge in RL-based solutions~\cite{wu2021dynstgat,xu2021hierarchically}: FRAP~\cite{zheng2019learning} models phase selection as a competition between phases, utilizing the intrinsic symmetry of intersections; CoLight~\cite{wei2019colight} introduces a graph neural network to model coordination between intersections. ~\cite{zhang2022expression}(Advanced-) and~\cite{wu2021efficient}(Efficient-) introduce more informative observations that improve the performance of existing methods. However, most existing methods address intersection heterogeneity by assigning each intersection an independent policy network, resulting in excessively high resource demands and low training efficiency when dealing with numerous city-scale intersections. Therefore, city-scale TSC necessitates the adoption of universal policy-based methods.

Constructing a \textbf{universal} policy for city-scale TSC has to address the heterogeneity of intersections, enabling the wide applicability of the policy to intersections with varying topologies and scales. Existing universal policy methods simplify this problem to tackling only the heterogeneous ego intersections' information, neglecting the effective utilization of their heterogeneous neighbor intersections. Specifically, MPLight~\cite{chen2020toward} designs universal observations of traffic states at intersections and uses parameter-sharing techniques to train large-scale solutions. UniTSA~\cite{wang2024unitsa} and GESA~\cite{jiang2024general} leverage the symmetry of intersections to improve the training efficiency of the universal policy. Attendlight~\cite{oroojlooy2020attendlight} and OAM~\cite{liang2022oam} design representation techniques that accommodate variable-length inputs, making the model applicable to intersections with varying topologies and scales. \textbf{However, the effective representation of the influence from neighbor intersections is largely unexplored, possibly due to the existence of intricate and varying traffic flow relations between intersections that challenge the construction of a universal and consistent neighbor influence representation.} Therefore, our CityLight aims to solve the challenge of consistently representing and aggregating neighbor influences, so that the universal TSC policy can learn shared coordinated policies that fit city-scale heterogeneous intersections.

\section{Preliminaries}
\subsection{Definitions}
\begin{figure}[t]
    \centering
    \includegraphics[width=0.97\columnwidth]{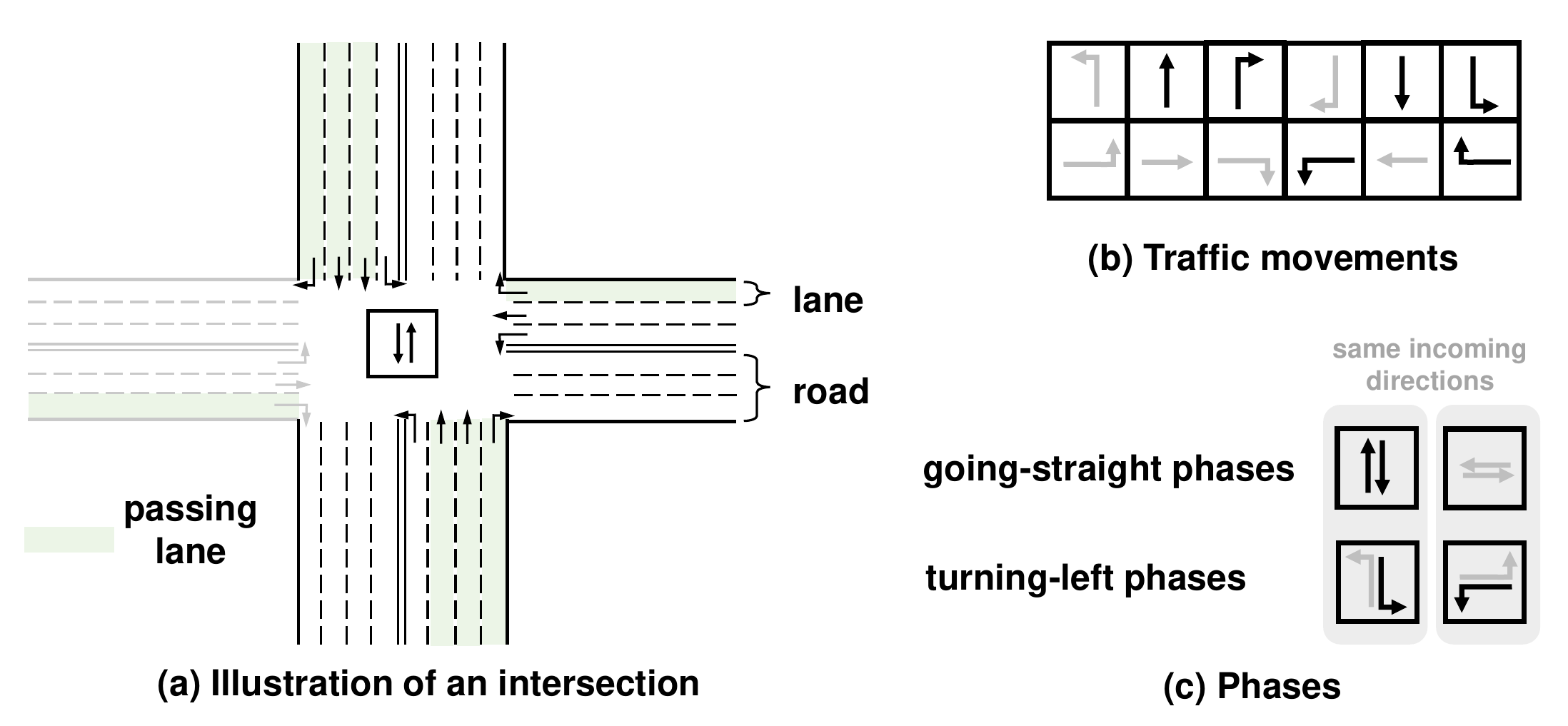}
    \caption{Illustration of (a) intersections, (b) traffic movements, and (c) phases. The grey area indicates the removed section for 3-arm intersections. We also annotate going-straight phases, turning-left phases, and phases from the same incoming direction in (c). Best viewed in colors.}
    \label{fig:definition}
    \Description{definition}
\end{figure}

\textbf{Road} (Fig.~\ref{fig:definition}(a)): Roads of an intersection are unidirectional edges connecting to or from the intersection. Each road has multiple lanes and each lane usually leads to one exact traffic movement at the intersection. 
\\
\textbf{Traffic Intersection} (Fig.~\ref{fig:definition}(a)): A traffic intersection is where multiple roads intersect and traffic signals are used to regulate the smooth flow of vehicles. Intersections in cities show \textbf{heterogeneity}, varying in topology (3-arm or 4-arm) and scale (number of connecting lanes).
\\
\textbf{Traffic Movement} (Fig.~\ref{fig:definition}(b)): The progression of vehicles across an intersection in a specific direction, namely, turning left, going straight, or turning right.
\\
\textbf{Phase} (Fig.~\ref{fig:definition}(a)): A set of two traffic movements that do not conflict and are generally allowed simultaneously. Consistent with the standard setting in traffic management and planning, the number of phases in the intersections can be three or four.
\\

\subsection{Task Setup}
Our research problem is a city-scale multi-agent coordination traffic signal control problem, with the number of agents scaling up to ten thousand levels. At each time step $t$, the partial observation $o_i^t$ of agent $i$, i.e. an intersection, includes the number of queueing vehicles of each phase, the number of lanes of each phase, and the passing state (which phase is the last passing phase). The objective of the task is to maximize the overall traffic efficiency of the system, which is commonly measured by the average vehicle travel time and overall vehicle throughput.

\subsection{Problem Formulation}
To apply RL training, we model the problem as a decentralized partially observable Markov decision process (Dec-POMDP). Dec-POMDP is parameterized by $(S,A,O,R,P,n,\gamma,h)$. $n$ is the number of agents. $S$ is the state 
space, $A$ is the joint action space. $o_i = O(s;i)$ is agent $i$'s observations at 
state $s$. $P(s' \mid s,a)$ defines the transition probability from state $s_i$ to state $s_i'$ via 
agents' joint action $a$. $R(s,a)$ is the reward function. $\gamma$ is the discount factor. The objective function is $J(\theta) = \mathbb{E}_{a,s}\left[\sum_t\gamma^t R(s^t,a^t)\right]$. In this task, all agents share the same policy $\pi_{\theta}$. The action space is the available phases at each decision-making step. 
          
\section{Method: CityLight}
\begin{figure*}
    \centering
    \includegraphics[width=1.95\columnwidth]{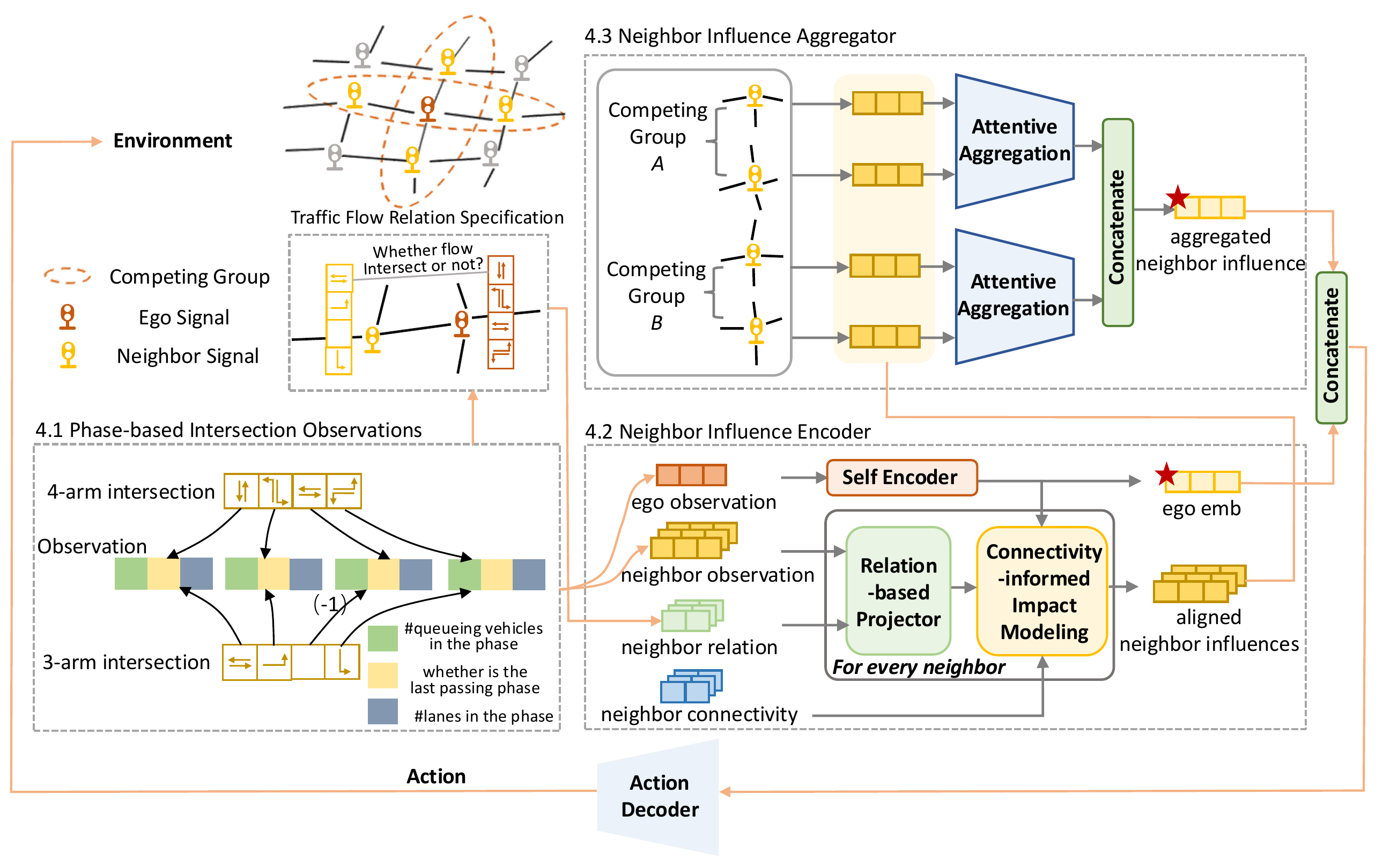}
    \caption{Illustrations of \textbf{CityLight}'s major modules for universally modeling and aggregating the influences of heterogeneous neighbor intersections. Best viewed in colors.}
    \Description{framework}
    \label{fig:framework}
\end{figure*}

Our proposed method, \textbf{CityLight}, aims to learn a universal coordinated TSC policy for large-scale heterogeneous intersections with diverse and intricate neighbor intersections, tackling the insufficient representation of neighbor influences in existing universal policy-based methods. The overall framework of CityLight is illustrated in Fig.~\ref{fig:framework}, which uses the centralized training with decentralized execution paradigm in MARL and assigns each agent a parameter-sharing policy to enable the learning of the universal policy. To obtain the the universal representations of heterogeneous intersections, CityLight first fixes the inner phase orders of heterogeneous intersections and constructs phase-based unified observations for them. Since the traffic flow relation space between adjacent intersections is narrowed after phase order fixing, an attention-based Neighbor Influence Encoder follows to model the influence of neighbors based on the specified neighbor relation and connectivity information. A Neighbor Influence Aggregator further attentively aggregates neighbor influences, accounting for mutual competitive relationships between neighbors. Inspired by mean field theory~\cite{yang2018mean}, we further design a neighborhood-incorporated reward to boost coordination in optimization.

\subsection{Phase-based Intersection Observation}
As real-world heterogeneous intersections consist of various numbers of lanes, it is hard to construct unified lane-level observations to represent the states inside intersections. Since the number of phases has a smaller range (only 3 or 4), we turn to construct the observation from the phase perspective. However, since the traffic flow relationships between the phases of neighboring intersections and the ego intersection are diverse (Fig.\ref{fig:background}(a)), observing phases in an arbitrary order creates a vast and complex traffic flow relation space between adjacent intersections, making it difficult to universally represent neighbor influences. Therefore, we propose a phase reindexing operation to fix the inner relative observation phase order for heterogeneous intersections. Specifically, we select any of the two \textit{going-straight phase} as the first phase, and choose the \textit{turning-left phase} with the same incoming direction as the second phase, followed by the other \textit{going-straight phase} and \textit{turning-left phase} pair as the third and fourth one (please refer to  Fig.~\ref{fig:definition}(c) for illustrations). A standard four-arm intersection can have two possible phase observation orders. \textbf{For each phase, we observe the number of queueing vehicles in this phase, its passing state (whether it is the last passing phase), and the number of lanes in the phase. The observation $o_i \in \mathbb{R}^{12}$ of an intersection $i$ is obtained by concatenating the phases' features along the order. For three-arm intersections, features of the missing phase are padded by -1 (Fig.~\ref{fig:framework}(a)). }

\subsection{Neighbor Influence Encoder}
\subsubsection{4.2.1 Traffic Flow Relation Specification}
To universally represent the influences of various neighbors, it is essential to specify the traffic flow relations of each phase of the neighboring intersections on that of the ego intersection. Since we have fixed the inner phase observation order, the traffic flow relation space between adjacent intersections is reduced from $4 \times 4$ to $2$, and can be classified by whether flows in their first phases intersect. Therefore, we specify this relation with a one-hot vector $S \in \mathbb{R} ^{1}$.

\subsubsection{4.2.2 Neighbor Influence Modeling Workflow}
Considering the validated effectiveness of the cross-attention mechanism in modeling the interactions between queries and values~\cite{vaswani2017attention}, we incorporate it to learn neighbors' relative traffic movement representations based on their observations and specified flow relation vectors. Specifically, with the relation vector $S$ as the query and the observation $o_j$ of the neighbor intersection $j$ as the value, the cross-attention mechanism adaptively maps the neighbor observation onto a relative traffic movement representation space. A self-attention layer further extracts features as the relative traffic movement representation $X_j^r$ of neighbor $j$ to ego intersection $i$, the process can be formulated as

\begin{equation}
   {X_j^r} = \mathtt{SA}(\mathtt{CA}( o_{j}, f(S_{ij}))), j\in \mathcal{M}(i).
   \label{equ:1}
\end{equation}
Here, $f \in \mathbb{R}^{1 \times k}$ is a linear layer, where $k$ is the embedding dimension and is a hyperparameter. $o_j$ is the observation of neighbor intersection $j$, and $S_{ij}$ is the specified relation of neighbor intersection $j$ to ego intersection $i$. $\mathcal{M}(i)$ is the set of one-hop neighbors, $\mathtt{SA}(r)$ and $\mathtt{CA}(r, q)$ are short for self-attention and cross-attention. 

Besides the traffic flow relation, the connectivity from neighbor intersections to the ego intersection, which can be indicated by distance and number of connected lanes from them, can also impact their influence intensities. Therefore, denoting the connectivity information with $C \in \mathbb{R}^2$, we again use cross-attention to model the connectivity-informed influence intensities. Specifically, using connectivity information $C$ and the ego intersection $i$'s feature $X_i$ as the query and the learned relative traffic movement representation $X^r$ of neighbors as the value, cross-attention outputs representations $X_{j}^n$ of the ego intersection's perceived influence. The process can be formulated as

\begin{equation}
   {X_{j}^n} = \mathtt{CA}\big(X_{j}^r, \mathtt{SA}(X_i)\oplus f(C_{ij})\big), j\in \mathcal{M}(i).
\label{equ:agg}
\end{equation}
Here $f \in \mathbb{R}^{2 \times k}$ is a linear layer, and $\oplus $ denotes vector concatenation. $\mathtt{SA}$ serves as a self encoder to extract informative parts from the ego intersection's observation $x_i$.

\begin{table*}[ht]
    \caption{The basic information and statistics of our five datasets.}
    \centering 
    \begin{tabular}{cccccc} 
        \toprule
        Datasets & Manhattan & Chaoyang & Central Beijing & Beijing & Jinan \\
        \midrule
        \#Intersections & 196 &97 &885 & 13952 & 4064\\
        \#Three-phase intersections & 0 &49 &630 & 9752 & 2466\\
        \#Four-phase intersections & 196 &48 &255 & 4200 & 1595\\
        \#Roads & 854 &608 &4640 &76806 & 23950\\
        Covered Area &25.3 $km^2$ &14.2 $km^2$ &79.3 $km^2$& 3104.4 $km^2$ & 1477.5 $km^2$\\
        \#Agents & 16317 &9000 & 55429 & 143298& 99712\\
        \bottomrule
    \end{tabular}
    \label{tbl:dataset}
\end{table*}

\noindent
\subsection{Neighbor Influence Aggregator}
Due to the rotational symmetry of intersections, the influence of reversed neighbors on the ego intersection is consistent. The two groups of reversed neighbors are typically arranged orthogonally, each exerting a competitive influence on the ego intersection. Therefore, directly concatenating or summing over the influence representations of multiple neighbors will overlook their mutual competitive relationship, resulting in insufficient representations of the overall neighborhood influences. To explicitly consider the inner correlations between neighbors, we regard each group of reversed neighbors as a competing group. For each group, we obtain its overall influence representations by attentively aggregating neighbors' influences, putting higher weight on noteworthy neighboring patterns. Denoting the competing group as $\mathcal{M}_{c_m}$, where $m$ denotes the competing group index and is $\in \{1, 2\}$, for neighbor intersections in $\mathcal{M}_{c_m}$, the process of calculating the attentive weights can be formulated as  

\begin{equation}
s_{j} =  \frac{e^{a_j}}{\sum\limits_{z \in  \mathcal{M}_{c_m}(i)} {e^{a_z}}}, j\in  \mathcal{M}_{c_m}(i), m\in \{0, 1\},
\end{equation}
where

\begin{equation}
 a_{j} = c\cdot \bm{{\rm Tanh}}(W\cdot X_{j}^n + b), j\in  \mathcal{M}_{c_m}(i), m\in \{0, 1\}.
\end{equation}
Here \( W \in \) $\mathbb{R}^{1 \times d}$, \( b \in \mathbb{R}^{1} \), and \( c \in \mathbb{R}^{1} \) are the learnable parameters, and \( d \) is the dimension of neighboring representations \( \{X^n\} \). The representation of each competing group is the weighted sum of its influence representations, and the overall representation of the ego intersection is the concatenation of its own representation with the representation of its two competing groups. Based on the representation, our action decoder, which is a multi-layer perception followed by a softmax layer, outputs the selection possibilities of candidate actions, i.e., the phases of the intersections.


\subsection{Neighborhood-Inclusive Reward Design}
Existing literature in traffic signal controls typically assigns independent rewards for intersections, aiming to maximize each intersection's traffic efficiency. For instance, releasing numerous vehicles from one upstream intersection to the already congested downstream intersection might enhance the traffic efficiency itself, but it could exacerbate extreme congestion at the downstream intersection, even leading to a gridlock. However, the combination of local optimal policies may not ensure a global optimal policy~\cite{xu2021hierarchically}. Thus, it is important to include neighbor intersections' situations in the optimization target to boost coordination further. Inspired by mean field theory~\cite{yang2018mean}, we design our reward as a weighted sum of the local traffic metric and the averaged neighborhood metric. Following established traffic RL works~\cite{wei2019colight,zheng2019learning,zang2020metalight}, we adopt the average queue length as the local metric $q_i$ for intersection $i$. The integrated reward $R_i$ is formulated as

\begin{equation}
R_i = q_i + \alpha \cdot \frac{\sum_{j \in  \mathcal{M}(i)} q_j}{|\mathcal{M}(i)|}.
\label{equ:reward}
\end{equation}
where $\mathcal{M}(i)$ denotes the one-hop neighbor set; $\alpha$ is a tunable coefficient to balance local-global optimization.

\section{Numerical Experiments}

\begin{table*}[t]
    \centering
    \caption{Average performances on three random seeds in five city-scale datasets. Bold denotes the best results and \underline{underline} denotes the second-best. '-' denotes 'no result' due to extremely high resource demands or expenses. TP and ATT are short for throughput and average travel time.}~\label{tbl:results}
\begin{tabular}{c |c c| c c| c c| c c| c c }
        \bottomrule
        & \multicolumn{2}{c|}{Manhattan} &  
        \multicolumn{2}{c|}{Chaoyang} & \multicolumn{2}{c|}{Central Beijing} & \multicolumn{2}{c|}{Jinan} &  \multicolumn{2}{c}{Beijing} \\
        \cline{2-3} \cline{4-5} \cline{6-7} \cline{8-9} \cline{10-11}
        Model &  TP$\uparrow$ & ATT$\downarrow$ & TP$\uparrow$ & ATT$\downarrow$ & TP$\uparrow$ & ATT$\downarrow$ & TP$\uparrow$ & ATT$\downarrow$ & TP$\uparrow$ & ATT$\downarrow$   \\
        \hline
        Fixed Time & $2343$ & $1687$ & $6076$ & $827$ & $18602$ & $1515$   & $33681$ & $1605$ & $60615$ & $1134$ \\
        Max Pressure & $2731$ & $1588$ & $1215$ & $1776$ & $5448$&$1880$&$2492$&$2052$&$15563$&$1567$\\
        Adjusted Max Pressure & $3371$ & $1499$ & $5909$ & $785$ & $\underline{18950}$ & $\underline{1497}$ &$33367$ & $\underline{1602}$ & $64613$ & $1090$ \\
        \hline
        LLMLight & $2325$& $1717$ & $1835$ & $1654$ & $7568$ & $1828$  & $-$& $-$ & $-$& $-$ \\
        \hline
        CoLight & $3480$ & $1519$ & $6661$ & $758$ & $18584$ & $1491$  & $-$ & $-$ & $-$ & $-$   \\
        Advanced-CoLight & $\underline{3523}$ & $\underline{1515}$ & $\underline{6730}$ & $\underline{749}$ & $17608$ & $1548$  & $-$ & $-$ & $-$ & $-$ \\
        \hline
        FRAP  & $1957$ & $1682$ & $6188$ & $920$ & $17156$ & $1576$  & $27358$ & $1715$ & $52720$ & $1112$  \\
        MPLight & $3136$ & $1575$ & $6104$ & $888$ & $17535$ & $1550$ & $30476$ & $1633$ & $60652$ & $1146$  \\
        Efficient-MPLight & $2819$ & $1636$ & $5916$ & $965$ & $16063$ & $1592$  & $29249$ & $1659$ & $57826$ & $1165$ \\
        Advanced-MPLight & $3339$ & $1548$ & $6347$ & $814$ & $16691$ & $1578$  & $\underline{33800}$ & $1609$ & $\underline{64764}$ & $\underline{1088}$ \\
        GESA & $2857$ & $1655$ & $6013$ & $955$ &$15748$ & $1608$ & $28199$ & $1698$ & $55814$& $1178$\\
        \hline
        \textbf{CityLight (Ours)} & $\textbf{3962}$ & $\textbf{1423}$ & $\textbf{7702}$ & $\textbf{600}$ & $\textbf{21548}$ & $\textbf{1440}$  & $\textbf{36742}$ & $\textbf{1554}$ & $\textbf{70451}$ & $\textbf{1030}$ \\
        \textbf{Improvement} & $\textbf{12.46\%}$ & $\textbf{6.07\%}$ & $\textbf{14.45\%}$ & $\textbf{23.65\%}$ & $\textbf{13.71\%}$ & $\textbf{3.97\%}$ & $\textbf{8.70\%}$ & $\textbf{3.01\%}$  & $\textbf{9.05\%}$ & $\textbf{5.34\%}$ \\
        \toprule 
    \end{tabular}
\end{table*}

\subsection{Experimental Setup}
This section summarizes the setting of our testbed, datasets, baselines, metrics, and implementations. 

\subsubsection{Datasets.} 
\begin{figure}[t]
    \centering
    \includegraphics[width=0.9\columnwidth]{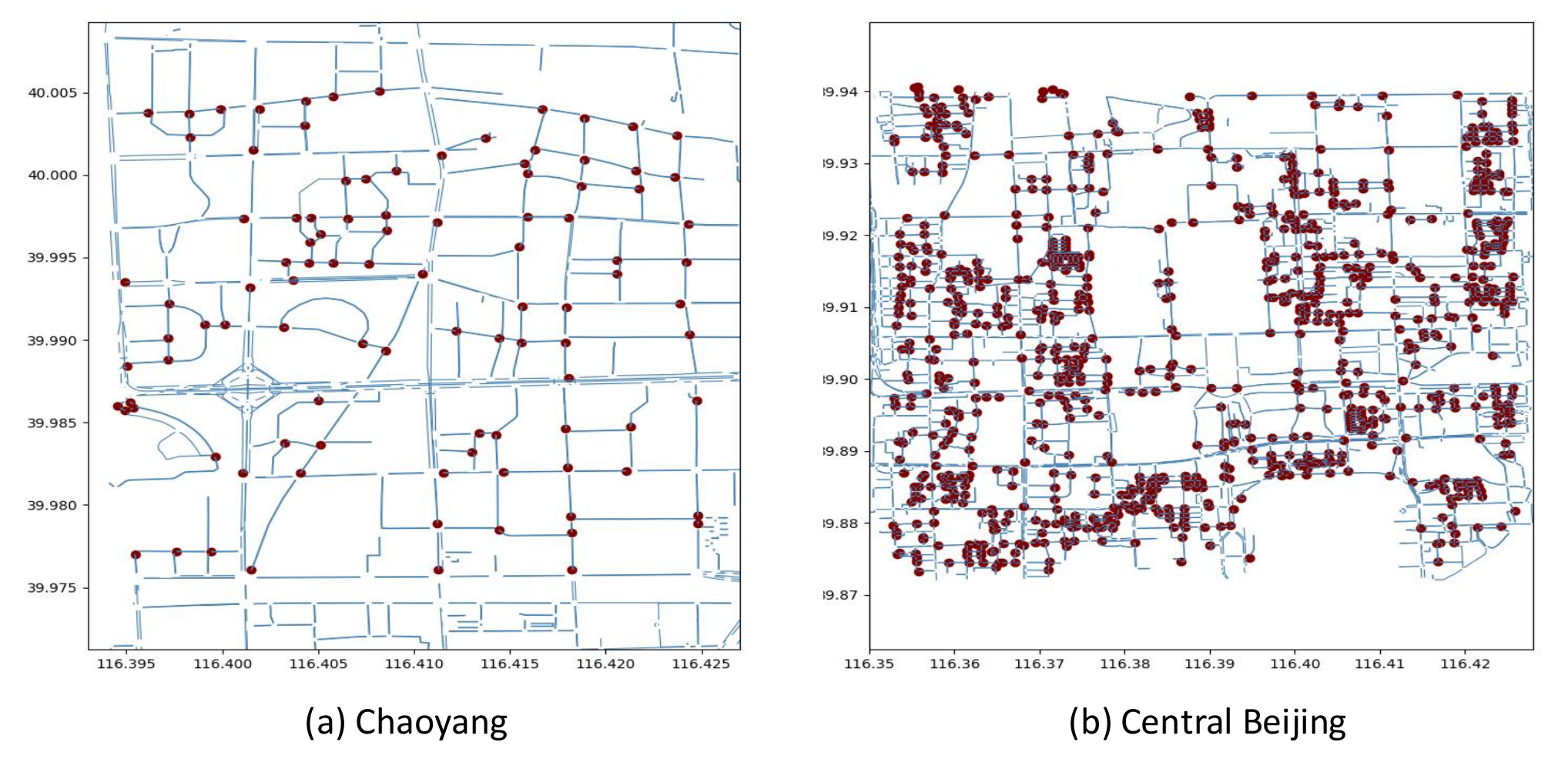}
    \caption{Illustrations of our collected real-world city-scale datasets: (a) Chaoyang; (b) Central Beijing.}
    \label{fig:dataset}
    \Description{dataset}
\end{figure}

Since we target city-scale TSC optimization, we incorporate one city-scale benchmark dataset, \textbf{Manhattan (196 intersections)}, and another four city-scale datasets collected from the real world: \textbf{Chaoyang (97 intersections)} (Fig.~\ref{fig:dataset}(a)), \textbf{Central Beijing (885 intersections)} (Fig.~\ref{fig:dataset}(b)), \textbf{Jinan (3930 intersections)} and \textbf{Beijing (13952 intersections)}. For the collected datasets, the road network is directly extracted from OpenStreetMap, and the vehicle flows are obtained from collected real-world traffic data: For Beijing, we obtain a 2020 hour-level OD flow matrix from a Chinese location-based service (LBS) provider; For Jinan, the traffic flows are recovered from traffic camera videos~\cite{yu2022spatio,yu2023city}. Statistics of the five datasets are presented in Table~\ref{tbl:dataset}.

\begin{figure}
    \centering
    \includegraphics[width=0.9\linewidth]{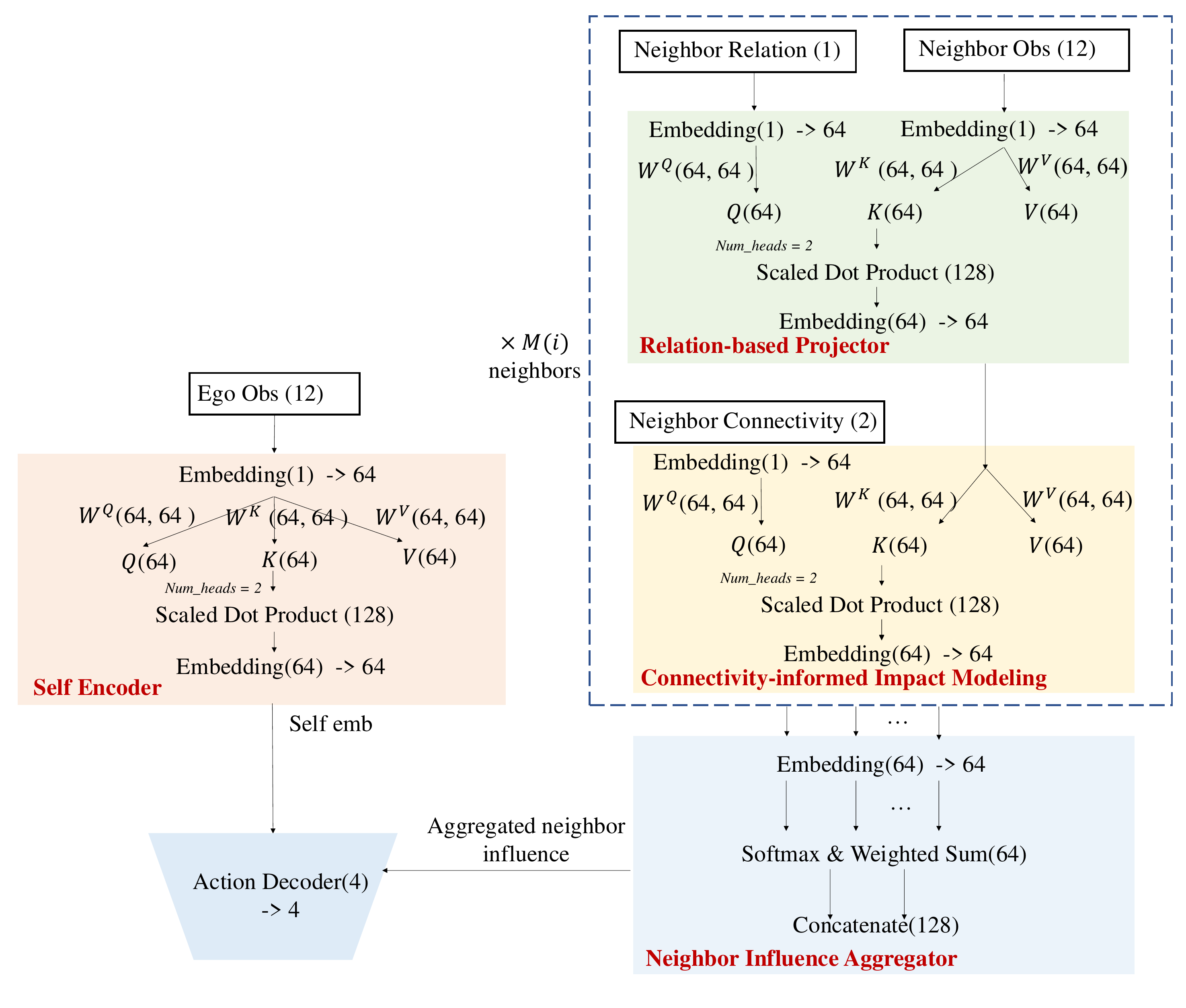}
    \caption{Architecture of CityLight.}
    \label{fig:architecture}
    \Description{architecture}
\end{figure}

\begin{figure*}[t]
    \centering
    \includegraphics[width=2\columnwidth]{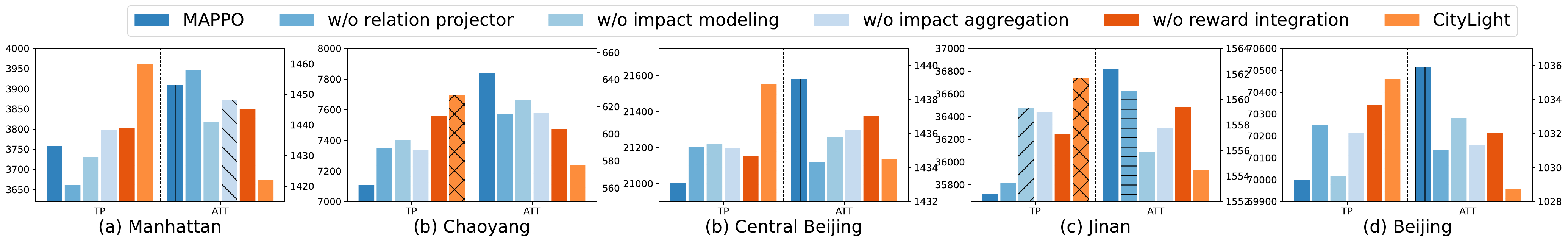}
    \caption{Performance of ablation variants.}
    \label{fig:ablation}
    \Description{ablation}
\end{figure*}

\subsubsection{Testbed.}
Since we target city-scale optimization, the simulation efficiency is vital for our testbed selection. Therefore, we refer to the \textbf{MO}bility \textbf{S}imulation \textbf{S}ystem (MOSS)~\cite{zhang2024moss}~\footnote{https://moss.fiblab.net/}. MOSS uses the well-acknowledged IDM car-following model~\cite{treiber2000congested} and the MOBIL lane-changing model~\cite{kesting2007general} to simulate the realistic motions of vehicles. Its simulation efficiency is achieved by the GPU acceleration, which has a 100 times acceleration compared to the existing CityFlow and SUMO simulators~\cite{zhang2024moss}. The experimental setting of the TSC follows existing work~\cite{wu2023transformerlight}: For a 60-minute timeframe, every 15-second time interval, each signal selects its next phase.

\subsubsection{Baselines and Metrics.}
For comprehensive comparisons, we compare with conventional rule-based, large language model-empowered, individual policy-based, and universal policy methods.  

\noindent
\textbf{Rule-based methods:}
\begin{itemize}[leftmargin=*]
    \item \textbf{Fixed Time~\cite{koonce2008traffic}.} Transit along the phase set with the fixed time interval.  
    \item \textbf{Max Pressure~\cite{varaiya2013max}.} Pressure is defined as the discrepancy in the number of vehicles for the entering lanes and existing lanes of a phase. Max pressure transits the traffic signal to the phase with the maximum value of pressure. In this way, the traffic system reaches equitable traffic flow. 
    \item \textbf{Adjusted Max Pressure~\cite{varaiya2013max}.} The length of roads in real-world road systems varies a lot, leading to wide-range vehicle capacity. We take into account the inherent discrepancies of road capacities and distill such impact by redefining pressure as the discrepancy in the number of vehicles per road length for the entering lanes and exiting lanes. 
\end{itemize}

\noindent
\textbf{Individual policy methods:}
\begin{itemize}[leftmargin=*]
    \item \textbf{CoLight~\cite{wei2019colight}.} A reinforcement learning-based method that uses a graph attentional network to model neighbor interactions. Each intersection has its own independent policy network. 
    \item \textbf{Advanced-CoLight~\cite{zhang2022expression}.} It adjusts CoLight by taking into account the running vehicles approaching the intersection in the observation. 
\end{itemize}

\noindent
\textbf{Universal policy methods:}
\begin{itemize}[leftmargin=*]
    \item \textbf{FRAP~\cite{zheng2019learning}.} A universal traffic signal control method that considers the rotation-symmetry of intersections by modeling from the perspective of phase competition relation. 
    \item \textbf{MPLight~\cite{chen2020toward}.} It adopts a parameter-sharing FRAP model as the unified policy network and introduces pressure of the intersection, which is the discrepancy in the number of vehicles for all entering lanes and all exiting lanes, as the reward.   

    \item \textbf{Efficient-MPLight~\cite{wu2021efficient}.} It introduces the efficient pressure into the observation space, which is calculated by the discrepancy in the number of vehicles per entering lane and per exiting lane of every phase.  
    \item \textbf{Advanced-MPLight~\cite{zhang2022expression}.} It adjusts MPLight by taking into account the running vehicles approaching the intersection in the observation. 
    \item \textbf{GESA~\cite{jiang2024general}.}A general scenario-agnostic (GESA) reinforcement learning framework for traffic signal control, which eliminates the need for heavy manual labeling, co-trains a generic agent across multiple city scenarios, and achieves superior transferability and performance in zero-shot settings.
\end{itemize}

\noindent
\textbf{Large language model-empowered method:}
\begin{itemize}[leftmargin=*]
    \item \textbf{LLMLight~\cite{lai2023large}.} It employs Large Language Models (LLMs) as decision-making agents for TSC. In our experiments, we use GPT-4o-mini as our agents. 
\end{itemize}

Following existing methods~\cite{zheng2019learning}, we introduce two common metrics for traffic efficiency measurement:

\begin{itemize}[leftmargin=*]
\item \textbf{Throughput (TP).} The total number of finished travels in the simulation episode. A higher throughput corresponds to a higher traffic efficiency.
\item \textbf{Average Travel Time (ATT).} The average travel time spent by all vehicles in the simulation episode. A low average travel time corresponds to a higher traffic efficiency. 
\end{itemize}

\subsection{Implementation}
Figure~\ref{fig:architecture} illustrates the overall architecture of our model, including detailed layer configurations and dimensions to facilitate reproducibility. During training, we adopted the Adam optimizer for gradient-based model optimization. The learning rates for both the actor and critic were set to 5e-4, following the default configuration of MAPPO~\cite{yu2022surprising}. We performed a grid search over other MAPPO-related hyperparameters, and the best configurations are available at \url{https://github.com/JinweiZzz/CityLight}. The balancing coefficient $\alpha$ was searched within the range {0.1, 0.2, 0.3}. For a fair comparison, we also fine-tuned the hyperparameters of all baselines on each dataset individually. All experiments were conducted on a single RTX 4090 GPU. The GPU memory usage of CityLight is approximately 20 GB. Training CityLight takes around 9 hours, which is comparable in efficiency to other universal policy methods and significantly faster than individual policy methods, which require over two days.

\subsection{Overall Performance}

\begin{table*}[t]
    \centering
    \caption{Generalizability of CityLight and three most competitive universal policy baselines across scale and city. Due to space limits, we shorten Chaoyang to 'CY' and Central Beijing to 'CB'.}
    \begin{tabular}{c c c c c|c c cc}
        \bottomrule
        & \multicolumn{2}{c}{$CB \implies CY$} & \multicolumn{2}{c|}{$CY \implies CB$} & \multicolumn{2}{c}{$Beijing \implies Jinan$} & \multicolumn{2}{c}{$Jinan \implies Beijing$} \\
        \cline{2-3} \cline{4-5} \cline{6-7} \cline{8-9}
        Model & TP$\uparrow$ & ATT$\downarrow$ & TP$\uparrow$ & ATT$\downarrow$ & TP$\uparrow$ & ATT$\downarrow$ & TP$\uparrow$ & ATT$\downarrow$\\
        \hline
        FRAP & $4988$ & $1055$ & $14522$ & $1667$ & $25190$ & $1704$ & $51920$ & $1304$\\
        MPLight & $5048$ & $1026$ & $15010$ & $1625$ & $26621$ & $1695$ & $52615$ & $1186$ \\
        Efficient-MPLight & $1955$ & $1626$ & $14966$ & $1626$ & $23144$ & $1749$ & $53175$ & $1208$ \\
        Advanced-MPLight & \underline{$5137$} & \underline{$926$} & \underline{$15040$} & \underline{$1623$} & \underline{$29933$}  & \underline{$1636$} & \underline{$57199$} & \underline{$1150$} \\
        GESA & $4691$ & $1092$ & $14409$ & $1702$ & $24880$ & $1721$ & $52310$ & $1270$\\
        \hline
        \textbf{CityLight (Ours)} & $\textbf{6704}$ & $\textbf{742}$ & $\textbf{19648}$ & $\textbf{1477}$  & $\textbf{34016}$ & $\textbf{1580}$ & $\textbf{66116}$ & $\textbf{1071}$ \\
        \textbf{Improvement} & $\textbf{30.50\%}$ & $\textbf{19.94\%}$ & $\textbf{30.64\%}$ & $\textbf{9.04\%}$  & $\textbf{13.64\%}$ & $\textbf{3.42\%}$ & $\textbf{15.59\%}$ & $\textbf{6.94\%}$ \\
        \toprule 
    \end{tabular}
    \label{tbl:transfer}
\end{table*}

Comparisons between CityLight and the 11 state-of-the-art baselines over five datasets are presented in Table~\ref{tbl:results}. As Colight and Advanced-Colight are individual policy methods that assign an independent policy network to each intersection, they demand excessively high memory and suffer from low training efficiency in city-scale optimization scenarios. LLMLight, which leverages large language models, also incurs prohibitively high API costs in such large-scale settings. Therefore, we do not apply these methods to the Jinan and Beijing datasets. Based on Table~\ref{tbl:results}, we have drawn these noteworthy observations:

\begin{itemize}[leftmargin=*]
    \item \textbf{Consistent Superiority.} Our CityLight consistently achieves the best performance across different metrics and datasets. Compared with the best baselines, the average performance gain of CityLight is 11.67\% in terms of throughput and 8.41\% in terms of average travel time. Such a consistent and large performance gain validates CityLight's effectiveness in dealing with city-scale intricate road networks and diverse traffic patterns.
    \item \textbf{CityLight outperforms existing non-universal policy-based coordination methods.} CoLight and Advanced-CoLight assign each intersection an independent policy network to learn coordination based on neighborhood representations aggregated with graph neural networks. In contrast, our approach consistently represents and aggregates influences from heterogeneous neighbor intersections with varying traffic flow interdependencies. This demonstrates that through training with diverse coordination scenarios across heterogeneous intersections, our learned universal strategy gains more effectiveness in tackling coordination. 
    \item \textbf{Incorporating neighbor information enhances the robustness and effectiveness of the learned universal policy.} As the datasets scale up, the heterogeneity of intersection topologies and traffic patterns increases, and the performance of baseline universal policy methods deteriorates, even falling below that of rule-based approaches. This demonstrates existing universal policy methods' applicability to diverse and intricate situations. In contrast, our CityLight significantly outperforms both rule-based and universal policy methods, showing its effective coordination and robustness of the learned universal policy.
\end{itemize}

\subsection{Generalizability Test}
Unlike individual policy methods that have to learn the individual policy networks from scratch when transferred to new regions, the universal policy trained in one region can be easily applied to others as it's commonly applicable to various intersection topologies. Therefore, we conduct comprehensive generalizability tests on universal policy methods, directly applying the learned policy in one region to the other. We select two transfer pairs: The first pair is Chaoyang and Central Beijing for evaluations of generalizability across scales. The second pair is Beijing and Jinan for evaluations of generalizability across cities. As shown in Table~\ref{tbl:transfer}, CityLight shows great transferability across both scale and city. The performance gain of CityLight over the best baseline even enlarges compared with the non-transfer scenarios, with an average improvement of 30.57\% when transferred across scales and 14.62\% when transferred across cities in terms of throughput. All these phenomena underline the robustness of the universal policy gained by uniformly considering neighborhood information. By learning from diverse neighborhood connectivity and traffic flow relations, the universal policy effectively captures a shared and widely applicable coordination strategy.


\subsection{Ablation Studies}
To assess the contribution of each component in CityLight, we conduct ablation studies on four key modules: the relation-based relative traffic movement projector, connectivity-informed influence modeling, competitiveness-aware influence aggregator, and neighborhood-inclusive reward design. As shown in Figure~\ref{fig:ablation}, removing the relative traffic movement projector reduces throughput by 2.23\%, highlighting the importance of consistently representing neighbor intersections. Excluding the connectivity-informed module leads to a 1.66\% drop, showing the value of modeling distance and lane connections. Removing the competitiveness-aware aggregator results in a 1.85\% decrease, confirming the benefit of capturing competitive and cooperative relationships between intersections. Lastly, adding the neighborhood-inclusive reward brings a 1.28\% gain, demonstrating its role in enhancing coordination. These results collectively validate the effectiveness of our components in building a universal coordinated TSC policy.

\subsection{Effectiveness of Our Universal Policy} 
In CityLight, we aim to develop a universal coordinated traffic signal control (TSC) strategy for heterogeneous intersections. To evaluate whether a single universal policy is more effective than training separate ones, we conduct two comparative experiments to see whether our universal representation can unify intersections with varying numbers of neighbors and generalize across regions with different road network characteristics, such as density and layout. The validation of these two aspects can further support the effectiveness of our method in consistently representing heterogeneous intersections and complex neighbor relationships, enabling a robust and generalizable universal policy. A theoretical analysis in Appendix A further supports the rationality of employing a 'universal' method, arguing that due to shared goals and patterns across intersections, the benefits of generalization may outweigh those of learning intersection-specific details.

\begin{table}[t]
    \centering
    \caption{Performance comparison of one universal policy across different topologies versus separate policies for each topology in terms of throughput.}~\label{tbl:config}
    \begin{tabular}{c c c c c c c}
        \toprule & Chaoyang &  
        Central Beijing & Jinan &  Beijing  \\
        \midrule
        Separate & $7325$ & $20944$ & $35943$  & $70031$ \\
        Universal & $7702$ & $21548$ & $36742$ & $70451$ \\
        \midrule
        \textbf{Improv.} &$\textbf{5.15\%}$  & $\textbf{2.88\%}$   &$\textbf{2.22\%}$& $\textbf{0.60\%}$ \\
        \bottomrule
    \end{tabular}
\end{table}


\begin{figure}[t]
    \centering
    \includegraphics[width=0.85\columnwidth]{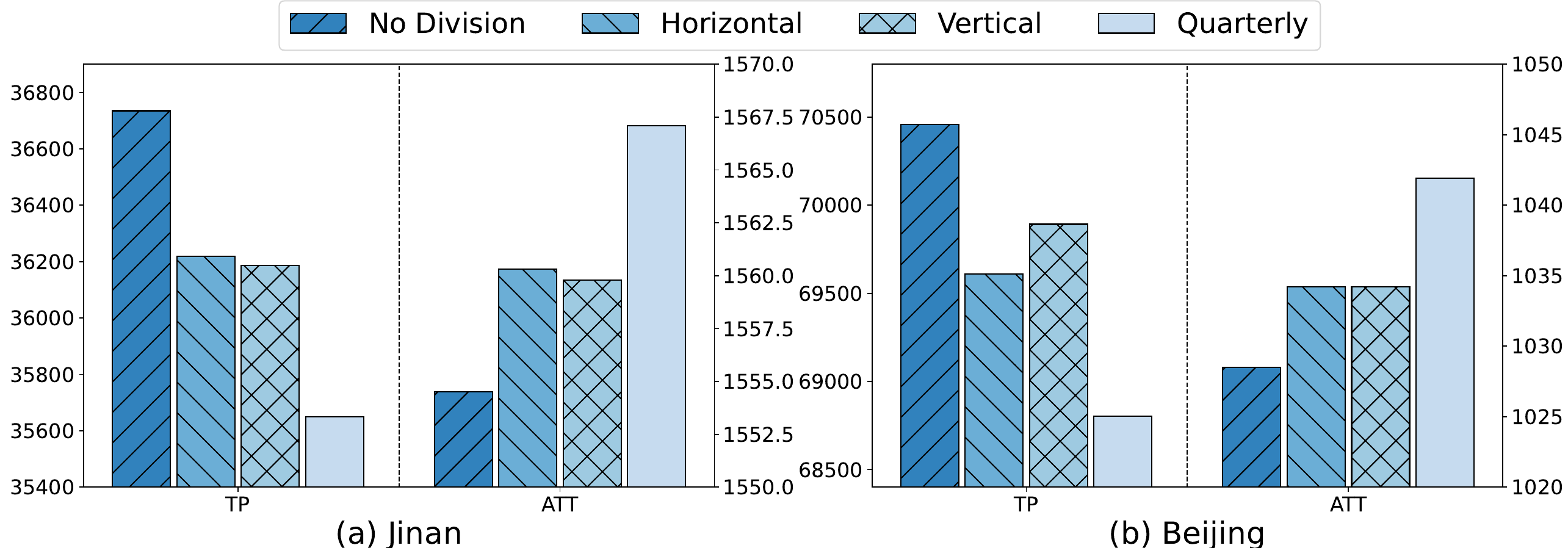}
    \caption{Performance comparison of one universal coordinated policy across space versus separate policies for regions divided horizontally, vertically, and quarterly in terms of throughput.}
    \label{fig:division}
    \Description{division}
\end{figure}


\subsubsection{Effectiveness of one universal policy across heterogeneous intersections.} As shown in Table~\ref{tbl:config}, training one universal policy for both three-arm and four-arm intersections together outperforms training separate policies respectively by $2.71\%$. Such a performance gain emphasizes our model's capacity to tackle heterogeneous intersections, which could stem from modeling the influence of neighbors and normalizing this influence through attentive fusion. As a result, the representation of neighbor influence will not be affected by the number of neighbors.

\subsubsection{Effectiveness of one universal policy across space.} Although conducting city-level traffic signal control is essential, it remains a question of whether intersections across space fit into a universal traffic signal control strategy. Thereby, we conduct a case study of CityLight in Jinan and Beijing where we divide the dataset into halves horizontally, vertically, and into quarters and learn the universal policy for easch subgroup respectively. As shown in Figure~\ref{fig:division}, as the number of sub-regions increases, the performance of the overall traffic signal control system declines. This phenomenon implies that while different regions of the city vary in terms of road network density, characteristics, and traffic patterns, our universal policy can learn their commonalities in traffic signal control by effectively aligning their representations.

\section{Conclusion}
In this work, we propose CityLight to learn a city-scale universal traffic signal control policy that achieves coordination. By consistently representing and aggregating neighbor influence out of intricate traffic flow relations and implicit competitive relationships among neighbors, CityLight learns a shared policy that can effectively incorporate neighborhood information for city-scale heterogeneous intersections. The great performance gain on five datasets that scale to the ten thousand intersections validates the effectiveness of CityLight. Further generalizability tests and case studies validate the wide applicability potentials of CityLight, underscoring its contribution to boosting coordination and enhancing traffic efficiency in city-scale traffic management.


\bibliographystyle{ACM-Reference-Format}
\bibliography{reference}

\clearpage

\appendix
\section{Theoretical Analyses}
In our method, we adopt the full parameter-sharing framework, MAPPO, to generate a universally applicable policy for heterogeneous intersections with intricate between-intersection interactions. Through our experiments, we found that such full parameter-sharing training is effective across different intersection configurations and the entire city, with superiority over assigning each configuration one parameter-sharing policy network and assigning each sub-region one parameter-sharing policy network. Here we present some theoretical analyses of how to explain such a phenomenon.  
\begin{proposition}[Full Parameter Sharing in TSC]
When controlling traffic signals in $N$ intersections with $N$ individual agents, full parameter sharing among all agents may be the optimal parameter-sharing strategy. 
\end{proposition}

Considering an arbitrary grouping strategy that divides the $N$ agents into $n$ groups, where the policy and state value function of group $i$ is denoted as $\pi^i$ and $V^i_{\pi^i}(s)$, respectively. We have:
\begin{equation}
    V^i_{\pi^i}(s)=\mathrm{E}_{a^i\sim\pi^i(a^i|s)}\left[r^i(s,a^i)+\gamma V^i_{\pi^i}(s')\right].
\end{equation}
The global state value function is the sum of all local state value functions as:
\begin{equation}
    \mathbf{V}_{\left\{\pi^1, ..., \pi^n \right\}}(s)=\sum_i V^i_{\pi^i}(s).
\end{equation}
We eliminate $(s)$ for short in the following analyses.

Considering a learning algorithm that approximates the optimal policy $\pi_*$ with $\pi_+$ regarding a given state value function $\Hat{V}_{\pi}$, where
\begin{equation}
    \pi_* = argmax_{\pi}\Hat{V}_{\pi}
\end{equation}
The algorithm's capability of approximation is denoted as:
\begin{equation}
    \Hat{V}_{\pi_*}-C_R\leq \Hat{V}_{\pi_+}\leq \Hat{V}_{\pi_*}-C_L,
\end{equation}
which holds for $\forall s$ with the constants
\begin{equation}
    0\leq C_L\leq C_R.
\end{equation}

When the parameters are globally shared among all agents, the learned policy
\begin{equation}
    \Bar{\pi}_+ = \Bar{\pi}^1_+ = ... = \Bar{\pi}^n_+
\end{equation}
and the optimal policy
\begin{equation}
    \Bar{\pi}_* = \Bar{\pi}^1_* = ... = \Bar{\pi}^n_* = argmax_{\pi}\mathbf{V}_{\left\{\pi, ..., \pi \right\}}
\end{equation}
satisfies
\begin{equation}
    \mathbf{V}_{\left\{\Bar{\pi}_*,...,\Bar{\pi}_*\right\}}-C_R\leq \mathbf{V}_{\left\{\Bar{\pi}_+,...,\Bar{\pi}_+\right\}}\leq \mathbf{V}_{\left\{\Bar{\pi}_*,...,\Bar{\pi}_*\right\}}-C_L,
\label{Eq.15}
\end{equation}
where
\begin{equation}
    \mathbf{V}_{\left\{\Bar{\pi}_*,...,\Bar{\pi}_*\right\}}=\sum_i V^i_{\Bar{\pi}_*}\quad
    \mathbf{V}_{\left\{\Bar{\pi}_+,...,\Bar{\pi}_+\right\}}=\sum_i V^i_{\Bar{\pi}_+}.
\end{equation}

When the parameters are locally shared among agents within each group, the learned policies
\begin{equation}
    \pi^1_+, ..., \pi^n_+
\end{equation}
and optimal policies
\begin{equation}
    \pi^i_* = argmax_{\pi}V^i_{\pi}
\end{equation}
satisfy
\begin{equation}
    V^i_{\pi^i_*}-C_R\leq V_{\pi^i_+}\leq V_{\pi^i_*}-C_L.
\label{Eq.19}
\end{equation}
Having
\begin{equation}
    \mathbf{V}_{\left\{\pi^1_*,...,\pi^n_*\right\}}=\sum_i V^i_{\pi^i_*}\quad
    \mathbf{V}_{\left\{\pi^1_+,...,\pi^n_+\right\}}=\sum_i V^i_{\pi^i_+},
\end{equation}
we can obtain
\begin{equation}
    \mathbf{V}_{\left\{\pi^1_*,...,\pi^n_*\right\}}-nC_R\leq \mathbf{V}_{\left\{\pi^1_+,...,\pi^n_+\right\}}\leq \mathbf{V}_{\left\{\pi^1_*,...,\pi^n_*\right\}}-nC_L
\label{Eq.21}
\end{equation}
by summing up Eq.~(\ref{Eq.19}) over $i=1,...,n$.

Traffic signal control for large-scale intersections is a relatively uniform task where the common goal is to maximize the overall traffic efficiency and there isn't mutual conflict and competence among intersections. Therefore, we may infer that some similarities exist for agents in traffic signal control. Considering the similarities among agents, the optimal policy with global parameter sharing, i.e., $\Bar{\pi}_*$, may approximate individual optimal policies, i.e., $\pi^i_*$ , as:
\begin{equation}
    V^i_{\pi^i_*}-L_R\leq V^i_{\Bar{\pi}_*}\leq V^i_{\pi^i_*}-L_L,
\label{Eq.22}
\end{equation}
which holds for $\forall s$ with the constants
\begin{equation}
    0\leq L_L\leq L_R.
\end{equation}

Combining Equ~(\ref{Eq.15}) (\ref{Eq.21}) (\ref{Eq.22}), we can obtain that when:
\begin{equation}
    L_L\leq C_L-\frac{1}{n}C_R,
\label{Eq.24}
\end{equation}
we have
\begin{equation}
    \mathbf{V}_{\left\{\pi^1_+,...,\pi^n_+\right\}} \leq\mathbf{V}_{\left\{\Bar{\pi}_+,...,\Bar{\pi}_+\right\}},
\end{equation}
indicating that the policy learned by full parameter-sharing among all agents outperforms policies learned locally within each group.

\section{Training Curves}
\begin{figure}[h]
    \centering
    \includegraphics[width=0.49\textwidth]{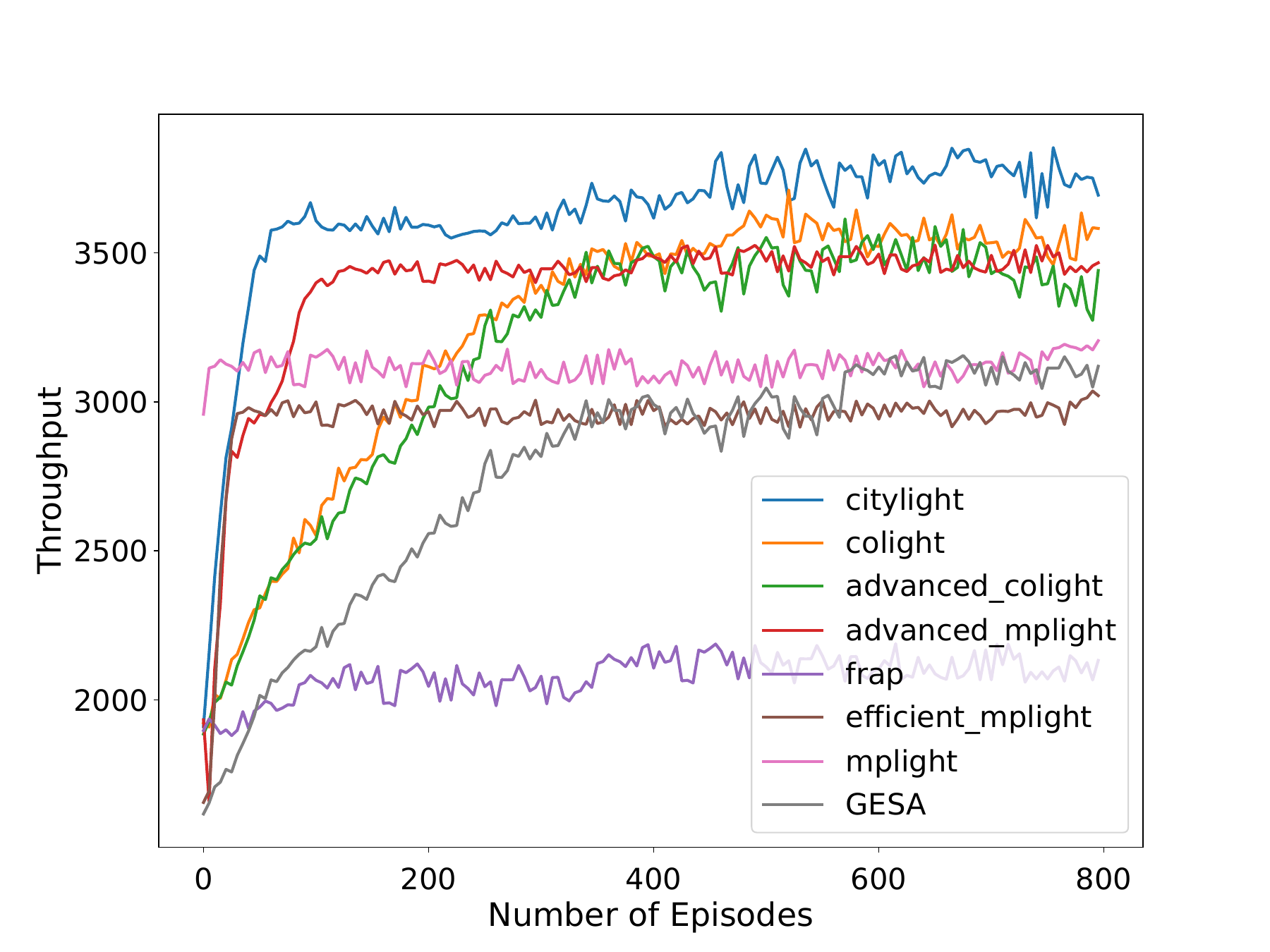}
    \caption{Training curve comparisons. Since the rewards of different methods are not uniform, we visualize the distribution of the throughput metric as the number of episodes increases.}
    \label{fig:curve}
    \Description{curve}
\end{figure}

We visualized the training curve of both our CityLight and the state-of-the-art baselines in Fig.~\ref{fig:curve}.   From the figure, we can see that all of the methods reach convergence. Parameter-sharing methods have a quicker convergence speed. CityLight not only has a relatively high convergence speed but also exhibits a stable and consistent superiority over all state-of-the-art methods.

\end{document}